\documentclass[manuscript]{acmart}
\setcopyright{none}
\AtBeginDocument{%
  \providecommand\BibTeX{{%
    \normalfont B\kern-0.5em{\scshape i\kern-0.25em b}\kern-0.8em\TeX}}}

\copyrightyear{2023}
\acmYear{2023}
\setcopyright{acmlicensed}\acmConference[EAAMO '23]{Equity and Access in Algorithms, Mechanisms, and Optimization}{October 30-November 1, 2023}{Boston, MA, USA}
\acmBooktitle{Equity and Access in Algorithms, Mechanisms, and Optimization (EAAMO '23), October 30-November 1, 2023, Boston, MA, USA}
\acmPrice{15.00}
\acmDOI{10.1145/3617694.3623235}
\acmISBN{979-8-4007-0381-2/23/11}

\usepackage{xcolor}
\usepackage{wrapfig}
\usepackage{multirow}
\usepackage{comment}
\usepackage{soul}

\usepackage{colortbl}
\usepackage{multirow}

\usepackage{subcaption}
\usepackage[labelformat=parens,labelsep=quad,skip=3pt]{caption}
\usepackage{graphicx}

\begin{document}

\title{Inclusive Portraits: Race-Aware Human-in-the-Loop Technology}

 \author{Claudia Flores-Saviaga}
 \email{floressaviaga.c@northeastern.edu}
 \affiliation{%
   \institution{Civic AI Lab, Northeastern University}
   \country{USA}
 }

 \author{Christopher Curtis}
 \affiliation{%
  \institution{Civic AI Lab, Northeastern University}
  \streetaddress{Boston, Massachusetts, USA}
  \country{USA}}

 \author{Saiph Savage}
 \affiliation{%
   \institution{Civic AI Lab, Northeastern University; }
   \institution{Universidad Nacional Autonoma de Mexico (UNAM)}
   \streetaddress{Northeastern University}
   \country{MX}}





\renewcommand{\shortauthors}{Flores-Saviaga, et al.}
\begin{abstract}
AI has revolutionized the processing of various services, including the automatic facial verification of people. Automated approaches have demonstrated their speed and efficiency in verifying a large volume of faces, but they can face challenges when processing content from certain communities, including communities of people of color. This challenge has prompted the adoption of "human-in-the-loop" (HITL) approaches, where human workers collaborate with the AI to minimize errors. However, most HITL approaches do not consider workers' individual characteristics and backgrounds. This paper proposes a new approach, called Inclusive Portraits (IP), that connects with social theories around race to design a racially-aware human-in-the-loop system. 
Our experiments have provided evidence that incorporating race into human-in-the-loop (HITL) systems for facial verification can significantly enhance performance, especially for services delivered to people of color.  Our findings also highlight the importance of considering individual worker characteristics in the design of HITL systems, rather than treating workers as a homogenous group. Our research has significant design implications for developing AI-enhanced services that are more inclusive and equitable. 
\end{abstract}




\begin{teaserfigure}
\centering
  \includegraphics[width=0.90\textwidth]{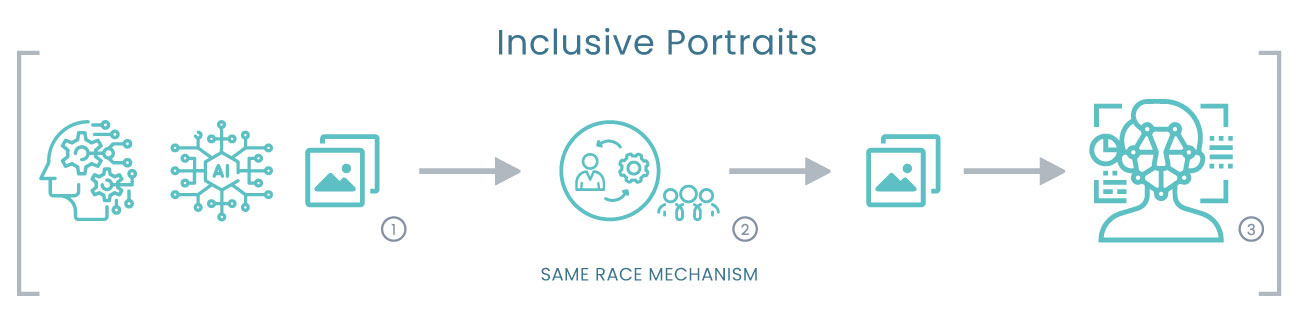}
  \caption{Overview of Inclusive Portait's functionality.}
  \label{fig:teaser}
\end{teaserfigure}


\maketitle

\section{Introduction}
Facial verification is the process by which a system confirms a person's identity via a facial biometric scan \cite{mayron2013secure}. Facial verification is a one-to-one mapping that responds to the question: {\it ``Is the person who they say they are?}'' Given its success in verifying a large number of people, facial verification has been integrated into a number of services. For example, several universities use facial verification for exam identity validation \cite{anand2022face}; banks and hospitals use facial verification to give people access to online banking or medical care \cite{bakunova2019biometrics}; governments use facial verification to give citizens access to protected buildings, or to classified government information \cite{jeon2019facial}. 

Current facial verification systems usually use AI (machine learning models) to verify people based on their facial features \cite{balaban2015deep,wang1804deep,wang2021deep}. AI-based facial recognition methods have gained popularity due to their ability to be easily deployed at scale, and their high accuracy in verifying faces \cite{huang2005face,o2007face}. Automated facial recognition methods are particularly effective when used under optimal lighting conditions, with frontal faces, and with high-quality images \cite{o2012comparing}. However, previous research demonstrated that past automated facial verification systems exhibited biases and errors \cite{raji2020saving,blanton2016comparison}. For instance, they might falsely identify people of color as criminal suspects at higher rates than white people \cite{buolamwini2017gender, huang2020face,zafeiriou2015survey}. In general, at least historically, automated approaches for verifying people can struggle to accurately recognize people of color \cite{perkowitz2021bias,fleischer2020bias}. One major factor is the lack of diversity in training data used to develop facial verification algorithms \cite{sham2023ethical}. When datasets do not include a representative sample of people with darker skin tones, the algorithms may struggle to correctly identify and verify these individuals \cite{roselli2019managing}. Additionally, some algorithms may be biased because they are trained on data that over-represents certain demographics or rely on features that are more prominent in certain groups, leading to inaccurate results for people of color \cite{daneshjou2021lack,mehrabi2021survey}. Finally, poor lighting and low-quality images can also impact the system's ability to detect and recognize facial features accurately \cite{jain2022imperfect,mehrabi2021survey}, which is usually more prevalent in images of people with darker skin tones \cite{buolamwini2017gender}.

One approach for mitigating the above concerns is technical: the creation of more accurate and more fair algorithms. Along these lines, the U.S. National Institute of Standards and Technology created a program for evaluating face recognition technologies~\cite{NIST:FRVT}, and indeed today's leading proprietary systems are significantly better than the past~\cite{ITIF:face}. In parallel, the limitations above have also led researchers to integrate human workers into the machine learning pipeline, to assist the automated methods in the facial verification process \cite{phillips2018face}. Humans may be better at recognizing other humans under a wide range of settings and can also recognize people as they evolve and change through time \cite{mohanty2019photo,mohanty2019second}. Facial verification systems may therefore improve further through the integration of human workers \cite{mohanty2020photo}. Such approaches are typically known as ``human-in-the-loop'' (HITL) \cite{wu2022survey}. Examples of HITL approaches within this context involve crowdworkers helping to verify unknown soldier portraits \cite{mohanty2020photo}, or detecting fraudulent passports \cite{white2015error}. While HITL systems can potentially mitigate some of the biases found in human-only or automated-only systems, they are still susceptible to biases and errors \cite{keswani2022designing}. Errors in HITL systems often arise because human workers may struggle to accurately identify people from different races \cite{wang2020mitigating}, particularly if they have limited exposure to facial expressions and features that are unique to those races \cite{yaros2019memory,o2005face}. The combination of inaccuracies and biases in facial verification systems has resulted in people of color being disproportionately affected \cite{buolamwini2017gender}. 
Hence, it is crucial to explore effective ways to integrate human and machine decision-making to address biases present in facial verification systems.

We believe that part of the problem is that most HITL systems either lack sufficient training of workers and/or usually overlook the diverse backgrounds of the human workers involved \cite{gordon2022jury,posada2022coloniality,alvarado2021decolonial,flores2020challenges}. To address the latter problem, ``jury learning" techniques have emerged \cite{gordon2022jury}. These techniques involve modeling workers and their backgrounds to better understand their perspectives and improve the quality of the labeling process \cite{gordon2022jury}. However, these techniques lack consideration for the racial backgrounds of workers, which can be a significant limitation \cite{kapania2023hunt}. Racial nuances can be essential to ensure that workers' unique experiences are accurately represented in their labor.

This paper introduces a novel racially-aware human-in-the-loop (HITL) system called "Inclusive Portraits" (IP). One of the key features of IP is its "Same Race" mechanism, which is based on social theory research related to race \cite{wong2020own,Lebrecht2009-ou}. This research suggests that, at least without training, people are more effective when working with data and information from their own race. The novelty of IP lies in its ability to leverage this insight to improve the accuracy and reliability of facial verification algorithms for people of color. By designing HITL systems that take into account individual worker characteristics, including race, IP can help to ensure that AI systems are more inclusive. We also discuss other approaches, such as more proactive training of workers, including inclusivity and diversity training, that are not considered here but could be opportunities for future study.

{\bf Context.}
%
%
 Before proceeding with our paper outline, we now step back and provide a broad, industry-level context for our research. Commercial face recognition systems have advanced significantly over recent years, with NIST's Face Recognition Vendor Test (FRVT) providing significant motivation for commercial advancement along with rigorous evaluations of the capabilities of commercial systems~\cite{NIST:FRVT}. Reflecting upon NIST's evaluations, leading proprietary systems have significantly fewer biases than past systems~\cite{ITIF:face}. That said, (1) not all entities using facial recognition technologies will have access to proprietary systems—some entities may build upon less accurate, openly available facial recognition technologies. Further, (2) biases do manifest in other types of technologies, beyond facial recognition systems. And, (3) even if current commercial face recognition technologies are better than past technologies, there is always value in further reducing any remaining biases in the overall systems that use those technologies. 

As we explore in this paper, human-in-the-loop systems (if designed well) have the potential to reduce remaining biases; the resulting system may have less bias than either machine-alone systems or human-alone systems. Our experiments are not with leading proprietary face recognition technologies, per the NIST FRVT results~\cite{NIST:FRVT}, but face recognition systems openly available to the research community. As such, we hope that our work is valuable to any entity building a composite system with openly available facial recognition technologies (1), that our results are valuable to future researchers and systems designers seeking to build systems leveraging other types of automated classification technologies (2), and that our results are valuable to entities with access to proprietary facial recognition technologies who wish to reduce any remaining biases even further.

{\bf Paper Outline.}
In the following sections, we provide an overview of research on bias in facial verification and discuss current work being done on human-in-the-loop (HITL) systems. We then introduce our novel racially-aware HITL system, Inclusive Portraits (IP), and describe its design. We then present the results of our evaluation of IP, which involved a study where crowdworkers used IP to verify the facial images of individuals from different racial groups. We compared the performance of IP with a state-of-the-art HITL approach and found that our racially-aware HITL system resulted in significantly higher accuracy in facial verification, especially for people of color. Finally, we discuss the implications and lessons that can be drawn from our research, including the importance of incorporating social theories of race into the design of AI systems as one path to ensure inclusivity.

{ \bf Positionality Statement}
As HCI researchers, we are committed to designing and evaluating technology that considers diverse perspectives and experiences. We recognize that facial verification technology can pose difficulties for people of color, especially when it is not designed with their needs and concerns in mind. Despite these challenges, we believe that this technology has the potential to be valuable in certain contexts and to benefit people of color. Thus, our research aims to create more inclusive technologies that address and combat biases against people of color. Our team consists of authors from diverse backgrounds, including two Latinas, one of whom holds a leadership role in their institution related to diversity, equity, and inclusion. Additionally, our academic work includes incorporating technology ethics and inequities into curricula.  
We acknowledge our own privilege and biases and strive to address these issues through a critical lens and by actively seeking diverse perspectives and experiences. Our goal is to create technology that is inclusive, accessible, and equitable for everyone. By contributing to the development of technology that truly serves and supports all communities, we hope to make a positive impact. We believe that by designing technology with a focus on inclusivity, we can help reduce the harm caused by biases and create a more just and equitable world.
\section{Related Work}
Literature related to our research includes three main pieces: (1) Facial Verification and Biases; (2) Human-in-the-loop systems; and (3) Race and Computing. 

\subsection{Facial Verification and Biases.}
In a recent survey, Wang et al \cite{wang2021deep} describe the problem of facial verification as the prominent biometric technique for identity authentication. Facial verification has been widely used in many areas, such as military, finance, public security, and daily life \cite{bennett2021s,aru2013facial,milligan1999facial,libby2021facial}. The field itself started to rise in popularity in the 1990s after the introduction of several breakthrough approaches, and the field continued to evolve, with deep learning methods taking the stage with aplomb in the early and mid-2010's \cite{al2019review}. Within the last decade the field has seen advances and the technology has become employed in law enforcement, social media, marketing, and more \cite{scheuerman2020we}. 

Face recognition is not a single type of technology, but rather a class of technologies. Face recognition can be divided into two types: face verification and face identification \cite{fuad2021recent}.  Face identification systems focus on identifying a single person out of a large set of people (1:N facial matching), while facial verification systems take two images and ask whether they are of the same person (1:1 facial matching). For both classes, it is important to consider the potential for bias. Indeed,  there is a real and prevalent concern regarding the bias in the datasets and models that are a part of learning systems that involve human actors and the consequences for the populations these AI-powered systems are applied to \cite{eubanks2018automating,benjamin2019race,hamidi2018gender}. One well-known bias is that some verification systems often have a higher identification accuracy for male faces than for female faces \cite{sixta2020fairface}. 
Scheuerman et al.\ also identify that in the case of human categories, race is an area of particular concern regarding bias in machine learning fairness literature \cite{scheuerman2020we}. Recent research showed that current machine-learning models have higher error rates when verifying the faces of US underrepresented populations, e.g., Asians or African-Americans \cite{kroll2022acm,raji2019actionable}. According to Phillips et al., \cite{phillips2011other}, like humans, even state-of-the-art face recognition algorithms --- or at least open, non-proprietary systems available for academic study --- struggle with ``other-race face'' recognition (see NIST's FRVT results for proprietary systems~\cite{NIST:FRVT}; the U.S. Department of Homeland Security has also tested proprietary algorithms~\cite{DHS:face}).
Benjamin observed that technologies used in other countries may manifest biases differently, e.g., algorithms ``developed in China, Japan, and South Korea recognized East Asian faces far more readily than Caucasians''~\cite[p.\ 112]{benjamin2019race}.
Cavazos et al. acknowledge that there could be different factors such as lack of algorithm training with representative faces,  inaccurately represented sub-populations, quality of the training, or test
photographs across subgroups \cite{cavazos2020accuracy}.
Currently, there are various research efforts to address forms of racial bias in the machine learning pipeline. Balayan et al.\ produced a recent survey of practical techniques \cite{10.1007/s00778-021-00671-8}. Scheuerman et al.\ explored how computer vision datasets are often at odds between social values and desired metrics as well as provided suggestions on how to reconcile these differences \cite{DBLP:journals/corr/abs-2108-04308}. Wang et al.\ proposed a reinforcement learning model which was successful in mitigating the racial bias across image datasets \cite{wang2020mitigating}. 

These efforts have highlighted the importance of addressing racial biases in machine learning and computer vision datasets. However, this research has tended to focus on the development of new machine learning models and techniques for quantitatively mitigating bias rather than on leveraging HCI principles and social theories for advancing new tools and dataset creation paradigms. Indeed, much previous research on this topic has failed to consider and take advantage of modern theories around race and computing; as Scheuerman et al.\ put it \textit{``despite increasing attempts to diversify databases, approaches remain simplistic and lacking in critical and social theories''} \cite{scheuerman2020we}. The contribution of this paper is in part to address this exact need. To enhance the accuracy of facial verification, especially for individuals of color, we integrate insights from HITL (human-in-the-loop) system research and race and computation research to create a race-aware HITL system. While our technical evaluations are focused on openly available, non-proprietary facial verification systems, we believe that all future automated systems can benefit from the thoughtful consideration of how to include humans in the loop.

\subsection{Bias in Crowdsourcing and Human-in-the-Loop Systems.}
Crowdsourcing and HITL systems face the challenge of human biases, which can affect the accuracy and fairness of AI models \cite{wang2019racial,wang2020mitigating,libby2021facial}. Researchers have proposed various solutions to address biases in crowdsourcing and HITL systems. These include the development of web applications that promote interpersonal accountability to mitigate biases \cite{thebault2019pairwise}, a web application influenced by cognitive psychology to help experts distinguish between labels and reduce biases \cite{mohanty2019second}, and modeling crowdworkers to predict their decision-making process, enabling designers of AI systems to choose the right crowdworkers to avoid potential biases \cite{gordon2022jury}. However, despite these efforts, there is still limited work on how to address biases in crowdsourcing and HITL systems, particularly biases that affect people of color. To create more accurate and unbiased outcomes, especially for these communities, this paper suggests incorporating ideas from race and computation into the design of HITL systems. By doing so, we can better understand how workers' different racial backgrounds can help to create solutions that are more inclusive and respectful of diverse perspectives. 

\subsection{Race and Computing.} Race is studied as a social construct \cite{bamshad2003does,baker1994role} --- racial groups are a product of social construct, not nature, and vary across social groups and time \cite{coates2004race}. In a computing context, prior work has adapted critical race theory for HCI, emphasizing that race is both real and socially constructed. It highlights how racial systems of oppression can be encoded and perpetuated in computing systems, such as racially biased facial recognition \cite{buolamwini2017gender}), and that engaging with race meaningfully in a design process is crucial to combating racial bias \cite{ogbonnaya2020critical,smith2021keepin}. In addition to critical race theory as a productive framing for understanding sociocultural systems, we can use it as a tool for design and evaluation of systems (e.g., by tracing lines of power and hierarchy, by drawing out how technical design choices interact with, reproduce, and/or mitigate racial inequities, by incorporating an understanding of how other identities co-mingle with race/ethnicity, etc.\ --- much in the same way that critical disability studies can be used to evaluate assistive technologies \cite{savage2022global, mankoff2010disability}). In this paper, we use theories around race and computation to design HITL systems that are better adapted to different races.

\section{Inclusive Portraits (IP)}
Inclusive Portraits (IP) is a novel Human-in-the-Loop (HITL) system that utilizes social theories related to race to boost the accuracy of facial verification, as illustrated in Figure \ref{fig:teaser}. The system has two main components: (1) an automated component, which receives pairs of facial images for verification and employs state-of-the-art machine learning techniques to complete the process; and (2) a human component, where crowdworkers come into play. For pairs of images where the automated component is uncertain, IP sends them to the human component for verification. This is where the system's uniqueness lies, as it introduces new ``Racially Aware Labor Interfaces'' that leverage the diversity of crowdworkers to improve the accuracy of facial verification. 


\subsection{IP's Automated Component.}
\label{sec:ip:automated}
The automated component of IP begins by receiving pairs of facial images from various datasets or sets of photos that require verification. These pairs are then sent to state-of-the-art facial verification models, as outlined in Table \ref{tab:study1AI}. It is worth noting that these AI models are all available for scientific study (excluding proprietary models, as noted in~\cite{NIST:FRVT}).

Using these models, IP calculates a ``confidence score'' for each pair of faces. The score represents IP's level of certainty that the faces belong to the same individual. To determine this score, IP takes the median confidence score of each AI model for the given pair of images, creating an overall measure of confidence. Pairs with lower confidence scores are compiled into a list and passed on to the human component for further analysis. This process allows IP to leverage the strengths of both automated and human analysis, ultimately improving the accuracy of the verification process.

{\bf Variations.}
In a production system, designers may instead aim to minimize false negatives or false positives, depending on the specific goals of their system. If minimizing false negatives is the priority, only uncertain negative outputs of the automated system may be sent to the human component. Conversely, if minimizing false positives is the priority, only uncertain positive outputs may be sent for human verification.

Another approach would be to send all negative or all positive outputs to the human component for further analysis. 
For instance, all negatives (regardless of how certain) could be sent to human verifiers,  who could then reclassify them as positives. This would lead to a system that reduces false negatives but increases false positives, resulting in a higher overall count of positives (all automated positives and all automated negatives with human positives).
Such an approach may come at the cost of increased resource consumption or processing time. Ultimately, the decision regarding which approach to use will depend on the specific needs and constraints of the system being developed.

\subsection{IP's Human Component.}
IP's {human component} incorporates a racially aware interface to efficiently manage the coordination of crowdworkers for the facial verification of images where the automated component is uncertain. Unlike traditional human-in-the-loop interfaces that assign ``anonymous'' crowdworkers to verification tasks \cite{mohanty2020photo}, IP's approach takes into account the diversity of workers' backgrounds and experiences. This results in a more robust and reliable verification process, especially when verifying the faces of people of color. By leveraging the diverse perspectives and expertise of the crowdworkers, IP's racially aware interface enables more accurate and inclusive verification of individuals.

\subsubsection{Racially Aware Interface.} 
Human-in-the-loop (HITL) interfaces often neglect the diverse backgrounds of the workers involved, which can result in AI systems that fail to capture the perspectives and experiences of certain groups  \cite{gordon2022jury,posada2022coloniality}. A promising approach to addressing this challenge is ``jury learning," which involves modeling workers and their backgrounds to improve the quality of the labeling process \cite{gordon2022jury}. However, a key limitation of this approach is its failure to account for workers' racial backgrounds. To overcome this limitation, IP introduces a racially aware interface that models workers based on their racial backgrounds. This innovative approach empowers workers to provide more tailored labeling that reflects the distinct needs and perspectives of their respective communities, including those of communities of color. The novelty of IP's racially aware interface lies in its unique annotation interface that incorporates a Same Race Mechanism. This novel mechanism is based on research that has shown that people are less effective when working with data and information from outside their own race \cite{wong2020own,Lebrecht2009-ou}. This component thus matches workers with labeling tasks related to their racial background \cite{rhodes2009race}. In specific, IP's racially aware interface matches workers with facial verification tasks that involve faces from their same racial background. This approach allows workers to leverage their unique experiences, resulting in more accurate and racially sensitive facial verifications. By incorporating workers' racial backgrounds in the task assignment process, IP ensures that the labeling process is more inclusive and representative of diverse communities. 

In line with the best practices that recognize race as a social construct \cite{smedley2005race}, IP follows established guidelines and requests first for crowdworkers to self-identify their racial identity. This approach is especially important in large-scale data collection efforts \cite{fang2019harmonizing}. However, to ensure consistency and accuracy, it is recommended that we then code these identities to pre-existing categories or labels. By adopting these methods, IP effectively accounts for the nuances of race \cite{chen2023collecting}.  IP utilizes a matching process that pairs a crowdworker's racial background with a corresponding facial verification task. This ensures that the crowdworker is verifying images that align with their background and experiences, ultimately improving the accuracy and inclusivity of the labeling process. IP leverages the APIs of popular crowdsourcing platforms like Toloka to source workers from geographic regions that have a diverse population from different racial backgrounds.

\section{Evaluation}
We study whether IP's racially aware interface can achieve higher accuracy in facial verification. We conducted an IRB-approved study to compare IP's accuracy with a state-of-the-art HITL system, in order to investigate this. The study took place from November 6th to November 15, 2022.

\subsection{METHODS}
To ensure the validity and reliability of our study, we meticulously recruited human participants and prepared the data, AI models, and HITL approaches used in our evaluation. Note that our participants were the crowdworkers in both the IP and control HITL conditions. Our data included facial photos from diverse racial backgrounds to be verified. The AI models were the first line of verification, sending any data for which the AI models were uncertain or had doubts to the HITL systems. These HITL systems included IP and the control HITL, which we utilized to compare and contrast our approach. Our comprehensive preparation ensured a thorough and accurate evaluation of our system's performance in facial verification.\\

\begin{figure}
\centering
\includegraphics[width=1.0\columnwidth]{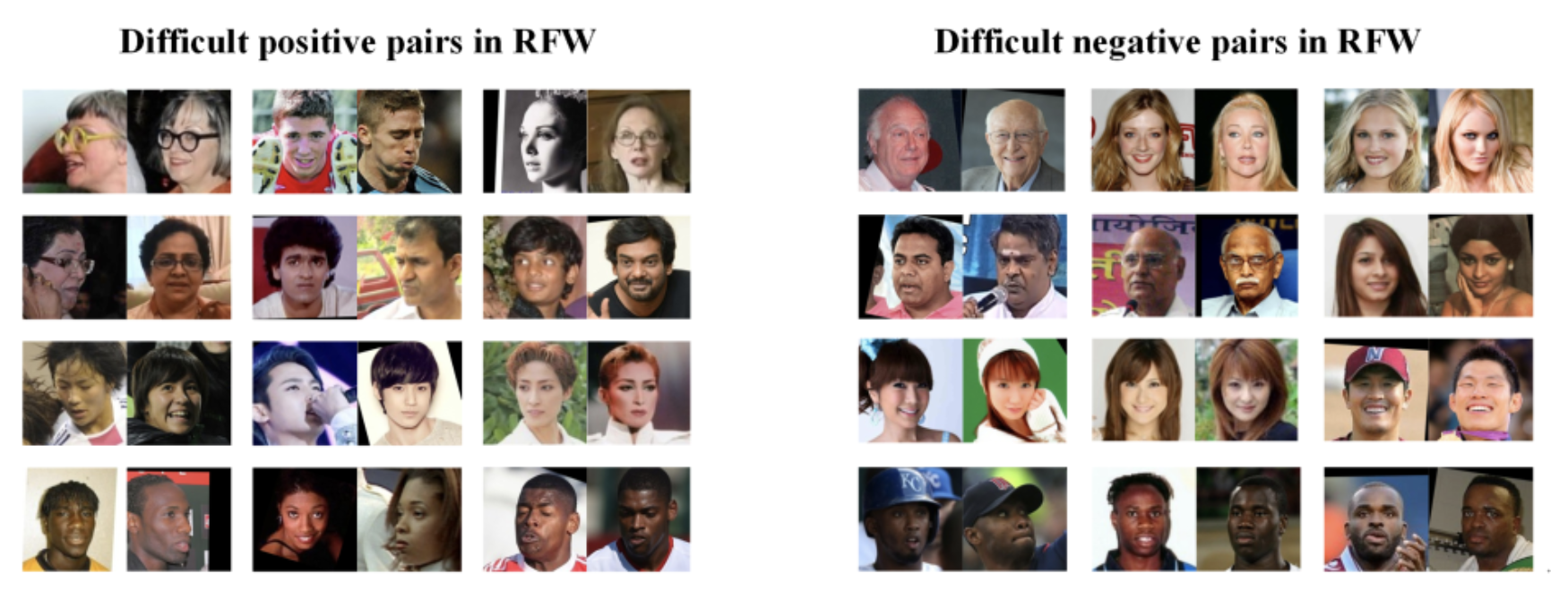}
    \caption{{Examples of images in our study, which included pairs of facial photos of people from different races. Note that across races, there were no  large differences in pose, age, or gender distribution in the facial images we used for our field experiments. }}
\label{fig:examples}
\end{figure}

{\bf Data}. To study and compare the effectiveness of IP and contrast it with a state-of-the-art HITL system, we utilized the ``Racial Faces in the Wild'' (RFW) balanced dataset \cite{Testing10:online}. This dataset contains 
images that are pre-labeled with the primary race present in each image. The RFW dataset is comprised of a 7K identities per race (Asian, Black or African-American, Caucasian and Indian (2,984 pairs of images).  Fig. \ref{fig:examples} shows examples of images from the RFW dataset. The RFW dataset includes both negative pairs (two pictures of different individuals) and positive pairs (two pictures of the same individual), making it possible to study the false positives and false negatives of a given approach. This is particularly important for uncovering biases in facial verification systems \cite{wang2019racial}. Furthermore, the RFW dataset stands out from other datasets because it provides a broad range of racial labels and was purposefully designed to facilitate investigations of biases in AI systems. By working with the RFW dataset and developing racially-aware HITL systems, we hope to contribute towards the development of additional approaches for more equitable and accurate facial verification technologies.

{\bf Participants (Crowdworkers).} A total of 160 participants were recruited for the study, evenly divided into two groups: 80 in the control HITL condition and 80 in the IP's condition. These participants were crowdworkers recruited through the Toloka crowdsourcing platform\footnote{\url{https://toloka.ai/tolokers}}.

In our study, it was essential to recruit crowdworkers from diverse racial backgrounds since we were comparing a racially aware HITL interface with a traditional HITL approach. To ensure unbiased and representative results, we aimed to have an equal number of crowdworkers from each racial group in both the control and IP conditions. Because race is a social construct \cite{benjamin2019race,park1950race}, we followed established guidelines and requested that crowdworkers self-identify their racial identity \cite{smedley2005race}. We then coded these identities into the labels used by the RFW dataset, which included the labels: ``Black or African American'', ``Asian'', ``Caucasian'', and ``Indian.'' This approach is important in large-scale data collection efforts \cite{fang2019harmonizing}, and helps to effectively account for the nuances of race \cite{chen2023collecting}. Additionally, we only selected crowdworkers who had no prior experience with facial verification tasks to control for confounding factors. Workers who were not selected were compensated for answering our recruitment survey.

\textit{\bf AI Models.} HITL systems in this domain of facial verification typically involve an AI component that performs the initial verification, followed by a human component that verifies and assists the AI with cases that the AI is uncertain about.\footnote{See earlier discussion of possible variations in HITL systems, e.g., systems that seek to maximize positives might send all negatives (and only negatives) to human verifiers.} Given this, we first prepared various AI models to perform automatic verification and then sent the images that all the AI models struggled with to our HITL systems (control and IP) for further verification. We evaluated eight open facial verification models (ArcFace, Facenet,  SFace, VGG-Face) and one proprietary model: Amazon Rekognition. We utilized the RFW dataset of facial photos and fed it to the AI models. Images that all the models struggled with were then sent to our HITL systems (control and IP) for verification. To ensure a fair comparison of how the HITL systems performed with different races, we made sure that the sample size of photos associated with each race was the same size across groups. In cases where a certain race had more errors, we randomly selected images until the smallest sample size was achieved, and we also followed best practices from HITL systems in terms of the number of images we gave to crowdworkers to verify \cite{mohanty2019second}.

We sent the HITL systems a sample of Asian (2000 pairs of images), Black or African-American (2000 pairs of images), Caucasian (2000 pairs of images), and Indian (2000 pairs of images) photos for verification. Note that both control and IP were given the exact same images for verification. 

Additionally, we considered it valuable to study the accuracy of each AI model to get a better sense of the information ecosystem present in this space.  To achieve this, we used the RFW testing dataset of facial photos and studied whether a given AI model correctly verified positive image pairs and rejected negative image pairs for each racial group. This allowed us to determine how well each AI model performed for a given race. It is worth highlighting that we did not utilize the RFW dataset to train the AI models that we tested. Our decision was based on the fact that training the models to reduce bias is beyond the scope of this research. Instead, we focused on studying how we could improve HITL systems by being racially aware.

{\bf IP \& Control HITL.} To compare IP with state of the art HITL approaches, we used a between-subjects study design and considered the following  conditions:
\begin{itemize}

    \item {\bf Control Condition}. In this condition, we recruited 80 crowdworkers from different races (African-American, Asian, Caucasian, and Indian), to perform face verification tasks using a traditional HITL interface.
    The participants were divided into four groups, each consisting of 20 workers from the same race. They verified a total of 32 images (8 images per race), sourced from the photos that the AI models were uncertain about. This condition served as the baseline for comparison with the Inclusive Portraits condition.  

     Note that to setup these conditions, we followed best practices from HITL systems in terms of the number of images we gave to crowdworkers. We limited the number of images each worker could verify to 32 to ensure they remained focused and attentive to the task at hand. This also helped to minimize any potential biases or fatigue that could arise from doing repetitive work \cite{ihl2018influence}, in this case verifying a large number of images. Additionally, we provided clear instructions and guidelines for the task, and we monitored the workers' performance to ensure the quality of the annotations. Overall, we followed standard best practices in HITL systems to ensure that our study was conducted in a rigorous and reliable manner.

     \item {\bf Inclusive Portraits Condition}. To make a fair comparison between the Inclusive Portraits (IP) system and the traditional HITL system, we recruited 80 crowdworkers, with 20 from each race, to perform facial verification tasks using IP. We ensured that the workers were assigned a task that involved verifying a similar number of images as in the control group. However, in this scenario, all the images belonged to the same race as the worker's self-identified race. The crowdworkers completed the verification process using the IP system, which was aware of their racial background. This enabled us to compare the performance of the racially aware IP system with the traditional HITL interface. 
    
\end{itemize}

\begin{wrapfigure}{r}{7.0cm}
\vspace{-1.0cm}
\centering
\includegraphics[width=.5\columnwidth]{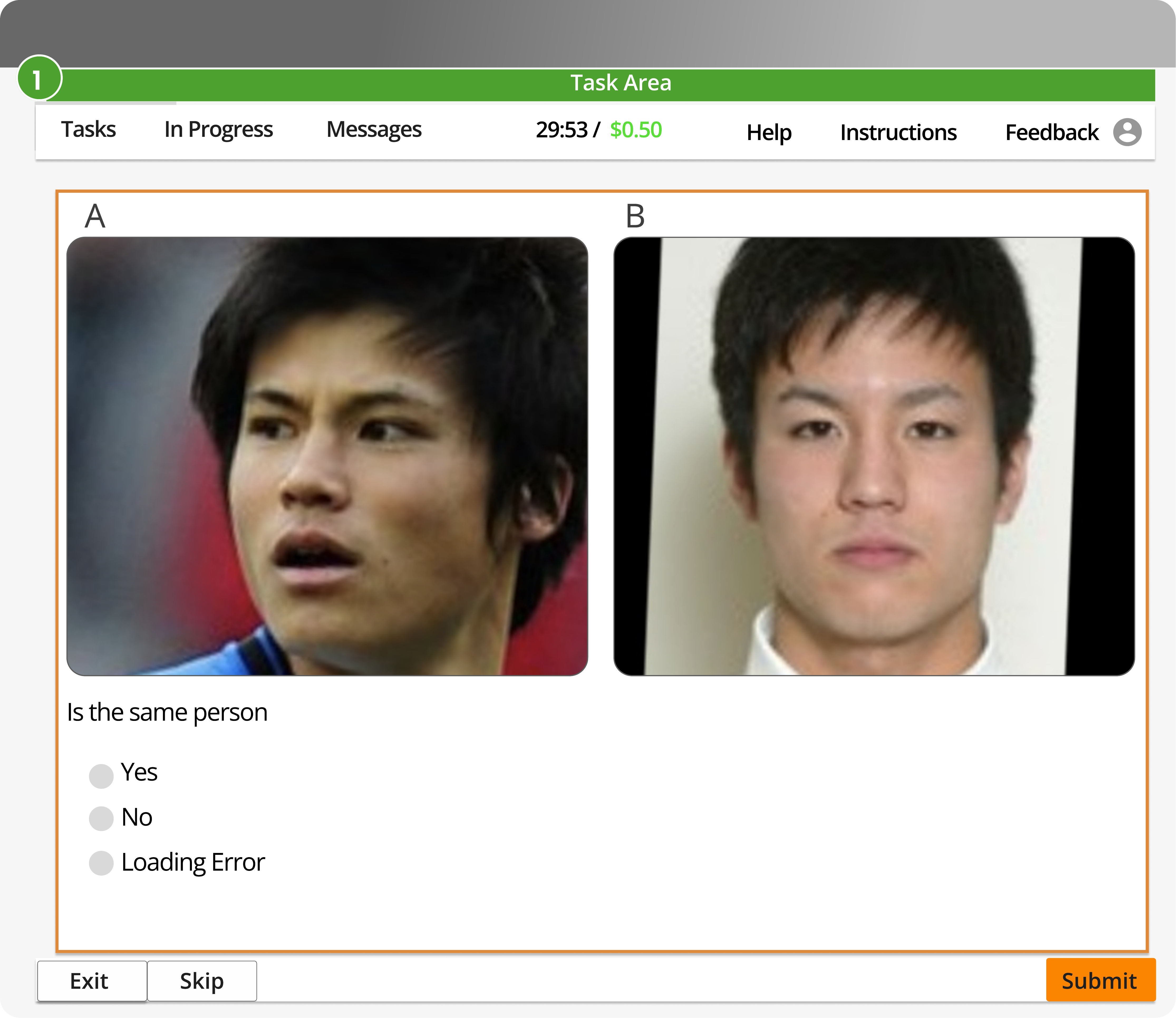}
 \vspace{-0.2cm}
    \caption{{Screenshots of the interface. The facial images are from the RFW dataset (described below), which allows for images to be included in papers.}}
    \vspace{-2.2cm}
\label{fig:screenshotInterface}

\end{wrapfigure} 

To prevent bias and ensure reliable data, both conditions used the same interface, see Figure \ref{fig:screenshotInterface}. Additionally, we implemented several measures to ensure that participants in our experiment had no prior exposure to the verification task. We meticulously screened participants to verify their lack of prior exposure to this task. Quality control questions were included to identify and exclude participants who were not fully engaged in the task. All workers passed these control questions. To compensate participants for their time and effort, we pay them \$10 USD, which exceeds the US minimum wage \cite{hara2018data}, as the entire experiment took 25 minutes.  

\subsection{RESULTS.}
Our study begins by presenting the accuracy of state-of-the-art AI models in verifying photos of people of color. We then highlight the limitations of these systems and proceed to compare a state-of-the-art HITL system for facial verification with IP's performance. It is important to note that the HITL systems were tested using the same data that the AI models struggled with, i.e., the photos in which they did not have confidence in their verification. 

{\bf{Results: AI Systems.}} 
To begin our study, we analyzed the accuracy of cutting-edge AI models in verifying faces from diverse races using the RWF dataset. To evaluate the performance of each AI model, we computed its accuracy for each race by dividing the number of correct predictions by the total number of images provided to the AI model for that specific race. As shown in Table \ref{tab:study1AI}, the AI models in general struggled with verifying faces of people of color. Specifically, ArcFace had an accuracy of only 25\% for Asian faces, while it had an accuracy of 80\% for Caucasian faces. This inconsistency was also observed in other AI models, such as Facenet, which performed poorly on Indian and African-American photos. Notably, Amazon's Recognition AI model had the best overall performance. However, it also exhibited a bias towards better accuracy on photos from Caucasian individuals.

These results are consistent with previous research findings \cite{wang2019racial}, and highlight the challenges that AI models still face when it comes to accurately verifying faces of individuals of color. 

\begin{table}[h]
\begin{tabular}{|c|c|c|c|c|c|}
\hline
\multicolumn{1}{|c|}{\textbf{\begin{tabular}[c]{@{}c@{}}Races\\(Photos)\end{tabular}}} &
  \multicolumn{1}{c|}{\textbf{ArcFace}} &
  \multicolumn{1}{c|}{\textbf{Facenet}} &
  \multicolumn{1}{c|}{\textbf{SFace}} &
  \multicolumn{1}{c|}{\textbf{VGG-Face}} &
  \multicolumn{1}{c|}{\textbf{Rekognition}} \\ \hline
\textbf{African-American}   & 30\%  & 45\%   & 75\%  & 75\% & 70\% \\ \hline
\rowcolor[HTML]{B3D7D7} 
\textbf{Asian}     & 25\%   & 75\%   & \cellcolor[HTML]{B3D7D7}75\% & 75\% & 70\% \\ \hline
\textbf{Caucasian} & 80\%   & 75\%   & 90\%  & 90\% & 100\% \\ \hline
\rowcolor[HTML]{B3D7D7} 
\textbf{Indian}    & 45\%   & 40\%   & \cellcolor[HTML]{B3D7D7}80\% & 80\% & 70\%\\ \hline
\textbf{Median (AI Model)}    & 30\%   & 45\%   & 75\% & 75\% & 80\%\\ \hline
\end{tabular}
\caption{Racial bias in open deep facial verification systems. Verification accuracy
(65\%) evaluated on 500 difficult pairs per race of RFW database are given. See NIST’s FRVT~\cite{NIST:FRVT} for an evaluation of proprietary facial verification systems; commercial deployments, which use facial verification technologies, may also have more advanced methods of curating image quality, which can impact results.}
\label{tab:study1AI}

\end{table}

\begin{figure}[h]
    \centering
    \includegraphics[width=0.5\textwidth]{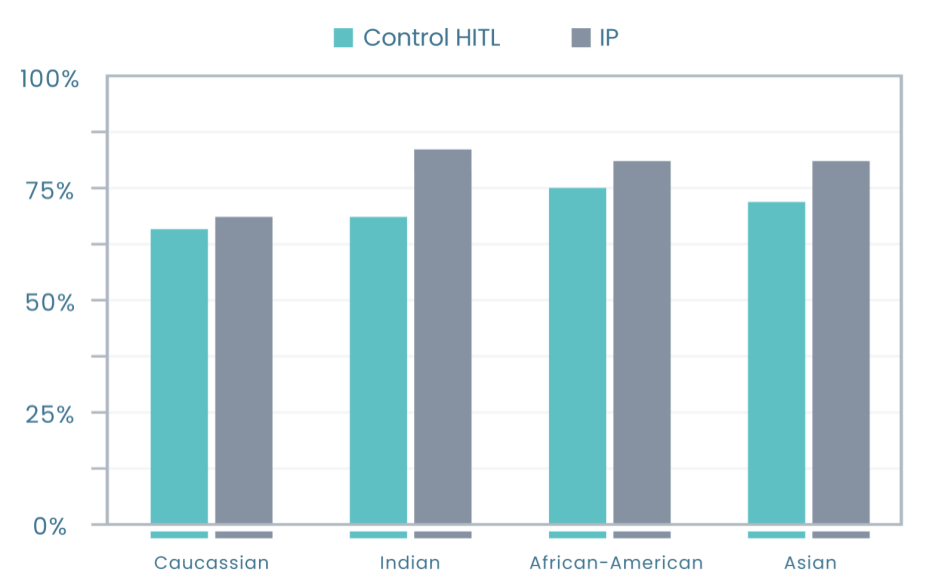}
    \caption{Overview of the median accuracy of the HITL control group and IP. The X-axis represents the different races of the crowdworkers performing the verification, while the Y-axis represents their accuracy.}
    \label{fig:chart}
\end{figure}

{\bf Results: HITL Systems.} To understand the performance of our control HITL condition and Inclusive Portraits (IP), we presented the images that were misclassified by (at least one of) the automated systems. In Fig. \ref{fig:chart}, we present a summary of the median accuracy of these two HITL systems across different races. Table \ref{tab:accuracy_std}, presents an overview of the accuracy and percentage difference in facial verification of different racial groups, using Control HITL and IP conditions.  

\begin{wraptable}{r}{5cm}

\begin{tabular}{|l|c|c|}
\hline
\textbf{Worker, Race} & \textbf{Statistic} & \textbf{p-value} \\ \hline
\rowcolor[HTML]{B3D7D7} 
\textbf{Caucassian}    & 1.3276             & 0.2492           \\
\textbf{Indian }       & 5.6366             & 0.0176           \\
\rowcolor[HTML]{B3D7D7} 
\textbf{African }      & 4.3395             & 0.0372           \\
\textbf{Asian  }       & 4.0060             & 0.0453           \\ \hline
\end{tabular}
\caption{Results of the Kruskal-Wallis Test comparing the accuracy of control and IP for face verification tasks.}
\label{tab:kw-table}
\vspace{-0.5cm}
\end{wraptable}

We observed that IP outperformed the state-of-the-art HITL system across races. To determine whether the differences in accuracy between IP and the control HITL system were statistically significant, we conducted a non-parametric analysis of variance. First, we tested the normality of the accuracy distributions in each condition using the Shapiro-Wilk test. The results showed that both distributions were non-normal, with p-values less than .05. We conducted the Kruskal-Wallis test, a non-parametric alternative to the one-way ANOVA, to study whether the differences in accuracy between these two condictions were significant, see Table \ref{tab:kw-table}.

\begin{table}
\begin{tabular}{|c|c|c|c|}
\hline
 &
  \textbf{Control HITL} &
  \multicolumn{1}{c|}{\textbf{IP}} &
  \multirow{2}{*}{\textbf{\begin{tabular}[c]{@{}l@{}}Percentage\\ Difference\end{tabular}}} \\ \cline{1-3}
\textbf{}       & \textbf{Accuracy} & \textbf{Accuracy} &           \\ \hline
\rowcolor[HTML]{B3D7D7} 
\textbf{African-American} &
  75.00\% &
  81.25\% &
  + 8.33\% \\ \hline
\textbf{Asian}           & 71.88\%           & 81.25\%           & +13.04\%  \\ \hline
\rowcolor[HTML]{B3D7D7} 
\textbf{Caucasian} & 65.63\%           & 68.75\%           & + 4.75\%  \\ \hline

\textbf{Indian}          & 68.75\%           & 84.38\%           & + 22.73\% \\ \hline

\end{tabular}
\caption{Overview of the accuracy and percentage difference in facial verification of different racial groups, using Control HITL and IP conditions.}
\label{tab:accuracy_std}

\end{table}

Our study demonstrates that there is a significant improvement in accuracy for people of color when using the IP system, as indicated by the positive percentage difference for each race. Specifically, the IP system resulted in a higher accuracy rate for Indian (+22.73\%), African American (+8.33\%), and Asian (+13.04\%) individuals, compared to the control system. The Kruskal-Wallis test statistics demonstrated that this accuracy improvement is statistically significant with values of 4.3395 (p-value 0.0372), 5.6366 (p-value 0.0176), and 4.0060 (p-value 0.0453), respectively. However, for the Caucasian race, there was no significant difference in accuracy between the control HITL condition and IP, as shown by the Kruskal-Wallis test statistics of 1.3276 (p-value 0.2492). 
This suggests that the IP system can effectively enhance accuracy in facial recognition, especially for people of color.

Our study highlights the potential benefits of using a racially aware HITL system like Inclusive Portraits for improving accuracy in facial verification tasks involving people of color. The fact that we did not observe an improvement in accuracy for the Caucasian race may be attributed to the existing bias towards this group in facial verification systems \cite{buolamwini2017gender}, resulting in a higher quality dataset and more experienced crowdworkers for this group. However, our results also emphasize the importance of having a diverse and inclusive workforce to mitigate bias in AI systems. By utilizing workers with different backgrounds and experiences, Inclusive Portraits was able to improve accuracy for people of color, demonstrating the potential of a more inclusive approach to HITL.

\section{Discussion}
In previous research, there have been attempts to combine human and machine intelligence to improve facial verification \cite{phillips2018face, mohanty2020photo}. Our study builds on this work by investigating how human-machine collaborations can be improved by taking into account workers' racial backgrounds. We introduce a racially aware HITL system, Inclusive Portraits, which significantly improves the accuracy of facial verification among untrained workers. By incorporating social theories on race into the design of human-in-the-loop systems \cite{sporer2001recognizing, wu2012through}, we can optimize and improve the outcomes of workers.

To further enhance the effectiveness of Inclusive Portraits, future work could explore the integration of training materials \cite{chiang2018crowd, dow2011shepherding, yu2014comparison}, combined with racially aware mechanisms to achieve higher quality labor \cite{Lebrecht2009-ou}. Our findings emphasize the value of considering the racial diversity of crowdworkers in facial verification tasks, especially in today's society where diversity and inclusion are increasingly valued. As improved facial verification technology can provide greater access to personalized services, Inclusive Portraits' racially aware interface is a valuable tool for enhancing the accuracy and quality of facial verification tasks.

\subsection{Design Implications.}
To expand the applicability of our study, we believe that future research should explore how architectures that consider workers' diverse backgrounds could be integrated into other human-in-the-loop systems, not limited to facial verification \cite{gordon2022jury}. The "Circuit of Culture" framework \cite{du2013doing, kimbell2011rethinking, hartley2012communication, hall1997representation} could serve as a useful tool for researchers to identify how workers' diverse backgrounds can influence the design of artifacts, especially in three key areas: (1) data representation, (2) data consumption, and (3) regulation.

{\it 1. Data Representation.} Representation is the creation of meaning through language and signs\cite{du2013doing,curtin2005privileging,bardzell2012critical}, which can be shaped by codes and symbols unique to specific communities, and are constantly evolving \cite{bardzell2012critical,tombleson2017rethinking}. We believe that designing human-in-the-loop (HITL) systems to coordinate crowdworkers to provide relevant data representations for end-users with similar backgrounds presents an opportunity for creating more inclusive and equitable AI systems. In natural language processing (NLP) and recommendation systems, HITL interfaces can be designed to coordinate crowdworkers from specific communities to provide translations, define relevant terms, or offer recommendations that are sensitive to their needs and perspectives. Incorporating the backgrounds of crowdworkers into the design of HITL systems has the potential to create more representative and community-driven AI solutions.

{\it 2. Data Consumption.} Hall's concept of the "decoding" of a message by its public, and the Circuit of Culture's idea of product meaning being shaped by consumption \cite{scherer2008cultural,hall1997representation}, both highlight the importance of involving end-users in the process of defining data representations. We believe there is value in bringing these ideas into HITL systems. Such an approach could help address biases and gaps in the definitions that crowdworkers' and designers' have, and the ones consumers have, and consequently create more representative labelled datasets \cite{du2013doing,bardzell2012critical}. However, conflicts may arise when consumers, crowdworkers, and designers have differing perspectives, so it is important to find ways to address these conflicts, as well as consider ways to prevent harmful or exploitative data representations, including through connecting with related cybersecurity research \cite{anwar2022human,samtani2020trailblazing}.

{\it 3. Regulation.} To ensure that the development and deployment of racially-aware technologies are done in an ethical and responsible way, it is essential to involve a diverse group of stakeholders, including members of the communities that these technologies are intended to serve, in the regulatory process \cite{du2013doing,bardzell2012critical,hall1997representation}. As previous researchers have noted \cite{flores2020challenges}, there is a need to explore designing crowdsourcing systems that take into account the backgrounds of crowdworkers. Taking into account crowdworkers backgrounds is important to provide input about the type of regulations needed in their community, ensuring that they are effective, equitable, and sensitive to the needs of the communities. Future research could also explore how to govern and design independent oversight bodies to effectively monitor the use of racially-aware technologies to ensure that these technologies are used in a responsible and ethical manner, while promoting social justice, respecting diversity, and protecting individual rights.

\subsubsection{Utilizing Own-Race Bias in Design Choices: Implications and Considerations.}  The own-race bias, a well-documented phenomenon where individuals tend to better remember faces of their own race than those of other races, has important implications for the design and implementation of facial verification systems \cite{wong2020own, Lebrecht2009-ou}. Our research has demonstrated that incorporating this knowledge into crowdsourcing platforms can lead to more effective design choices. However, it is important to carefully consider the ethical and logistical implications of federating and pooling workers by their race. Prior research on cross-race bias suggests that training methods can be used to mitigate this bias and reduce the disadvantage faced by outgroup workers \cite{tanaka2009neural, Lebrecht2009-ou}. Employers could use this knowledge to provide training to specific groups of workers who can then act as an auditing or control group or be used when there are not enough task workers to fill desired groups. Nonetheless, it is important to recognize that such strategies may not fully eliminate the own-race bias but rather aim to minimize its impact. Therefore, organizations must continually monitor the performance of their systems and make adjustments as necessary.

\subsubsection{Crowdworker Privacy and Racially Aware HITL Systems.} 
It is essential to consider worker privacy when designing and deploying racially-aware human-in-the-loop systems. The collection and processing of sensitive personal information, including race, can raise significant privacy concerns for workers involved in such systems. In many cases, workers may not be aware that they are contributing to the development of racially-aware systems or how their personal information is being used. Moreover, workers may face risks such as stigmatization or discrimination based on their participation in such systems. To mitigate these concerns, it is important to implement clear and transparent policies around the collection, storage, and use of worker data but also respect a user’s expectations for how their data will be used \cite{redmiles2021need}. For example, workers should be informed about the nature of the task, how their data will be used, and the steps being taken to protect their privacy. Additionally, organizations should implement strict data security measures and limit access to personal information to authorized personnel only. Further, anonymity and pseudonymity can be employed as ways to reduce the potential for workers' personal information to be associated with their work  \cite{kobsa2003privacy}. Finally, it is important to establish mechanisms for workers to report privacy violations or concerns and provide a means for workers to opt-out of the system if they choose to do so. By considering worker privacy concerns and implementing appropriate measures, organizations can help ensure that racially-aware human-in-the-loop systems are developed and deployed in a responsible and ethical manner.

{\bf Limitations and Future work}
Although our research sheds light on the effects of own-race bias on facial verification, it is important to acknowledge its limitations and the need for future work. In some settings, the race of individuals may not be known and, in such cases, it may 
not be possible or appropriate to infer race algorithmically; indeed algorithmic approaches for race detection may also exhibit biases \cite{wood2017instrumentation}. 
Further, the people being verified may be multiracial and the worker pool may not include people with the exact same demographic backgrounds. 
Additionally, exploring the impact of various worker backgrounds, such as age \cite{megreya2015developmental}, gender \cite{megreya2011sex},  personality \cite{megreya2013individual} and expertise \cite{flores2021fighting,flores2016leadwise,flores2022datavoidant}, on the effectiveness of facial verification and other types of tasks would provide valuable insights. Additionally, examining the role of monetary incentives on crowdworkers' performance in facial verification tasks could also be beneficial, as it has been shown to affect task quality in other contexts \cite{flores2020understanding,rogstadius2011assessment}. A comprehensive understanding of how these factors interact with own-race bias can facilitate the development of more equitable and accurate facial verification systems. Therefore, our research serves as a crucial initial step in creating more inclusive and fair systems, not just for face recognition systems, where industry and government continue to advance technically~\cite{NIST:FRVT}, but for AI systems as a whole. Future work may also study the role of workers in different types of classification systems as surveyed
in Section~\ref{sec:ip:automated}, e.g., 
a system designed to minimize false negatives at the expense of a greater number of false positives.



\section{Conclusion}

%
%
Humans and technologies each have their own strengths and their own weaknesses, and by combining both via human-in-the-loop systems it is possible to leverage the best of both and produce more equitable results. This paper explores how to push the human-in-the-loop frontier further and, in particular, explores the impact of different methods for involving humans in the loop.

Specifically, we study both a standard and an advanced way of including humans in a composite human-in-the-loop verification system. We experimented with our approach using openly available facial recognition technologies, which are an example of an automated technology with known biases (we did not experiment with the most advanced, proprietary facial recognition technologies according to NIST’s FRVT~\cite{NIST:FRVT}). Although leveraging open systems means that the underlying facial recognition technologies we studied do not have the same performance as leading proprietary systems, leveraging these openly available systems allowed us to explore human-in-the-loop composition methods using resources generally available to the academic community and hence resources that other researchers can build upon. Through our research, we developed Inclusive Portraits, a race-aware human-in-the-loop system that outperforms a state-of-the-art HITL system. Our system is grounded in social theory research on race, and our online study demonstrated its efficacy, particularly in the verification of non-Caucasian faces. Given our findings, we encourage future researchers studying and building advanced human-in-the-loop systems to consider fully the impacts and biases of individual humans in human-in-the-loop systems.\\

{\bf Acknowledgements.} This work was partially supported by NSF grant FW-HTF-19541. 
\bibliographystyle{ACM-Reference-Format}
\bibliography{sample-base}

\appendix

\end{document}